%% file: main.tex
\documentclass[journal]{IEEEtran}
\usepackage{cite}
\usepackage{amsmath,amssymb,amsfonts}
\usepackage{algorithmic}
\usepackage{graphicx}
\usepackage{textcomp}
\usepackage{xcolor}
\usepackage[hyphens]{url}
\usepackage{booktabs}
\usepackage{hyperref}
\usepackage{xspace}

\title{Why Do Prefetchers Fail? Let Agents Answer.}
\author{
  Xiangfeng Sun\textsuperscript{$\dagger$},
  Ceyu Xu\textsuperscript{$\dagger$,*},
  Ningzhi Ai,
  Zeyu Zhu,
  Yiyang Yuan, and
  Yuan Xie \\[1ex]
  The Hong Kong University of Science and Technology,
  Hong Kong SAR, China \\[0.5ex]
  \textit{xsunbv@connect.ust.hk, eeentropy@ust.hk,
  ningzhi.ai@connect.ust.hk,} \\
  \textit{eezeyuzhux@ust.hk, sasasasatori@gmail.com,
  yuanxie@ust.hk}%
  \thanks{\textsuperscript{$\dagger$}Equal contribution.}
  \thanks{\textsuperscript{*}Corresponding author.}
}

\newcommand{\alectobaseline}{Alecto\xspace}
\newcommand{\alectofinal}{MoP\xspace}
\newcommand{\alectofull}{Mixture of Prefetchers (MoP)\xspace}

\begin{document}
\maketitle

\begin{abstract}
Hardware prefetchers are crucial to modern processor performance, yet their design
remains a tedious, labor-intensive, and expert-driven process. Architects inspect
program execution and memory-access traces, identify recurring patterns, translate
them into online hardware heuristics, and implement and evaluate these heuristics in
simulation, often with no guarantee of improvement. Over decades, this process has
produced hundreds of mechanisms, but the pattern space remains far from exhausted:
human experts cannot systematically inspect billion-instruction traces across the
diversity of real-world workloads.

AI agents offer an opportunity to scale prefetcher autoresearch. Prior agent-driven
approaches, however, primarily rely on a model's parametric knowledge or
human-specified design spaces. Without being driven by measured failures, they lack a
scalable way to decide what capability to add next. We present a
performance-anomaly-driven autoresearch flow that repeatedly asks why a deployed
prefetcher fails and uses the diagnoses to construct the \textbf{\alectofull}. Each
iteration localizes high-impact unexplained misses to program counters, provides agents
with hardware logs, source code, and sliced execution traces, validates diagnoses
through runnable minimal cases, and synthesizes specialized sub-prefetchers for
recurring pattern families. Measured performance and remaining anomalies provide
feedback and context for subsequent iterations, enabling simulator-in-the-loop
discovery beyond model priors.

The autoresearch campaign consumes 1.91 billion DeepSeek V4 Pro tokens. Evaluated on
SPEC CPU2006 and SPEC CPU2017, the resulting prefetcher achieves a 61.1\% geomean IPC
speedup over no prefetching, outperforming three state-of-the-art human-designed
prefetchers---Alecto, Berti, and Pythia---by 14.5\%, 21.6\%, and 23.6\%, respectively.
RTL synthesis in a 6nm library reports 110\,KB of on-chip storage and
0.0347\,mm$^2$ area. To our knowledge, this is the first empirical demonstration that
an agent-driven hardware-design process can produce an RTL-practical prefetcher that
outperforms state-of-the-art human designs on unseen workloads.
\end{abstract}

\input{src/intro_human_closed_loop}
\input{src/background_llm}

\input{src/background_prefetching}

\input{src/flow}
\input{src/scaling}

\input{src/methodology}
\input{src/evaluation}

\input{src/conclusion}

\bibliographystyle{IEEEtran}
\bibliography{refs}

\end{document}

%% file: src/intro_human_closed_loop.tex
\section{Introduction}
\label{sec:intro}

\begin{figure}[!t]
  \centering
  \includegraphics[width=\columnwidth]{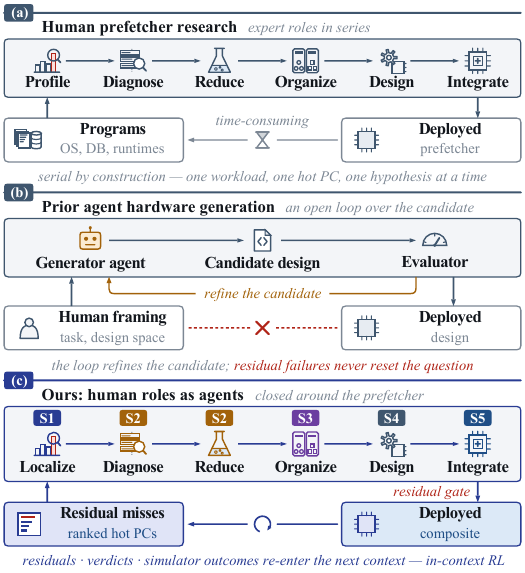}
  \caption{Learning the organization and closing the loop. (a) Human prefetcher
  research is effective because engineers divide the work into evidence-producing
  roles, from profiling and diagnosis to reduction, design, and integration, but the
  serial workflow does not scale. (b) Prior agent hardware generation usually begins
  from a human-specified task or fixed design space. Evaluator feedback can refine a
  candidate, but it does not turn the deployed design's residual failures into the next
  research question. (c) We instantiate the human roles as specialized agents and
  close them around the deployed prefetcher. Residuals, experimental verdicts, and
  simulator outcomes re-enter the next iteration's context, while residual gating lets
  accepted repairs accumulate without damaging existing coverage.}
  \label{fig:motiv}
\end{figure}

Hardware prefetchers try to hide long memory latency by predicting which data a program
will use and fetching it before the processor asks for it. The design of hardware
prefetchers conventionally follows this paradigm: when a deployed prefetcher still
leaves performance on the table, engineers first measure the workload to find where the
processor continues to wait on memory. They then inspect the instructions responsible
for those misses, compare low-level hardware observations with source code and execution
traces, and construct a small test that preserves the same behavior. Only after the
failure is understood do they propose an online mechanism and evaluate the integrated
design in simulation. Figure~\ref{fig:motiv}a summarizes this paradigm. Repeated over
decades, it has produced a rich catalog of stride and stream
~\cite{chen1995effective,jouppi1990stream}, spatial
~\cite{somogyi2006sms,bakhshalipour2019bingo}, temporal
~\cite{jain2013isb,grannaes2011dcpt}, and online-learning-enabled
prefetchers~\cite{bera2021pythia,hashemi2018learning}. Each design targets a particular
memory-access pattern, but real programs combine many patterns and shift among them over
time, so every prefetcher still leaves important misses uncovered.

The fundamental problem is scale. This human-driven paradigm asks a small number of
experts to search a pattern space created by an enormous and continuously changing body
of software. Real processors execute operating systems, databases, language runtimes,
services, scientific applications, and other codebases that collectively contain
billions of lines of code, under inputs and execution contexts that create still more
memory behavior. The resulting traces produced across the world's processors are far
beyond what any engineering team could systematically inspect. Human analysis proceeds
one workload, one hot instruction, and one hypothesis at a time, while the supply of
software and traces grows much faster than expert attention. Indeed, even fixed and
carefully curated suites such as SPEC CPU2006 and SPEC CPU2017 have been studied for
years without exhausting their useful prefetching patterns. The seemingly endless set
of undiscovered memory-access patterns is therefore not evidence that each pattern is
impossible to understand; it is evidence that manual discovery explores only a tiny
fraction of the available behavior.

Recent agent-driven design work attacks a different bottleneck: generation. Language
models assist RTL and EDA tasks~\cite{liu2023chipnemo}, generate Verilog
~\cite{liu2023verilogeval,thakur2023verilog}, and explore hardware from
natural-language specifications~\cite{fu2023gpt4aigchip}; evolutionary agents improve
programs against automated fitness functions
~\cite{romera2024funsearch,novikov2025alphaevolve,openevolve2025}. These systems show
that an imperfect generator can become useful when candidates are filtered by a strong
evaluator. However, as Figure~\ref{fig:motiv}b highlights, generation usually begins
\emph{after} a human supplies the task, guessed mechanism, fixed design space, or scalar
objective. Candidate feedback can improve an answer within that framing, but it does
not reveal which capability a deployed architecture lacks, distinguish absent
prediction from inaccuracy or lateness, or determine how a repair can coexist with
mechanisms already working. The agent automates artifact production while the human
still performs the architectural research that decides what should be produced.

We ask whether agents can instead learn the \emph{workflow and organization} of human
engineers, and operate inside a closed feedback loop with the deployed artifact. Our
answer is a performance-anomaly-driven autoresearch flow, illustrated in
Figure~\ref{fig:motiv}c. We preserve the human division of labor as a set of specialized
agent roles with explicit evidence contracts, then connect those roles to a
cycle-accurate simulator and the residual misses of the current prefetcher. The model is
not asked to guess a prefetcher from an empty prompt. It is asked to explain a measured
failure, reproduce the explanation, design for the reproduced family, and observe what
changed after integration.

This feedback acts as a form of \emph{in-context reinforcement learning} over the design
trajectory. By this term we do not mean that the agents update their model weights.
Instead, every iteration changes both their context and their environment: residual
miss locations, accuracy and lateness signals, minimal-case verdicts, compile and
simulator outcomes, and accepted mechanisms become inputs to subsequent work. A
successful engine removes its pattern from the residual; the next iteration therefore
sees a harder and more specific problem. A failed diagnosis returns a negative
experimental result rather than becoming design folklore. Unlike a fixed fitness loop,
this process updates the research question as well as the candidate answer.

Concretely, the flow begins from the strong three-engine Alecto
ensemble~\cite{li2025alecto}. S1 simulates the deployed composite and ranks hot residual
miss program counters. S2 assigns one analysis agent to each program counter with three
aligned views, namely its hardware log, source code, and sliced trace, and requires a
runnable minimal case. S3 groups reproduced causes into pattern families and emits a
design contract. S4 uses evolutionary code-generation agents to propose, compile, and
simulate one sub-prefetcher for each family. S5 integrates the surviving engine and
remeasures the composite, closing the training loop; only after the loop converges does
S6 evaluate the final design on held-out traces that took no part in research.

Closing the feedback loop also requires improvements to accumulate. If an agent edits a
strong prefetcher monolithically, large edits destroy existing coverage while small
edits merely retune it. Independently generated engines can likewise interfere through
cache pollution, bandwidth, and miss-status resources. We therefore append each repair
under \emph{residual gating}: every sub-prefetcher trains on all accesses but may issue
only when earlier engines issued nothing. A shared sandbox suppresses duplicates, and
a ground-truth accuracy throttle disables the extension layer when precision collapses.
An accepted engine becomes part of the environment seen by the next iteration without
rewriting the engines that preceded it.

The resulting campaign consumes 1.91 billion DeepSeek V4 Pro tokens and discovers
fourteen specialists on top of Alecto's three base engines. Design uses 30 training
traces, while 11 memory-intensive SPEC CPU2017 traces remain unseen until final
evaluation. On these held-out workloads, the resulting \textbf{\alectofinal} achieves
a 61.1\% geomean IPC speedup over no prefetching and outperforms Alecto, Berti, and
Pythia by 14.5\%, 21.6\%, and 23.6\%, respectively. Successive integrations raise or
preserve held-out geomean IPC. The complete 17-engine prefetcher uses 110\,KB of state
and synthesizes to 0.0347\,mm$^2$ in a 6nm library, showing that the closed agent
organization produces a practical microarchitecture rather than an unconstrained
simulation artifact.

\input{src/aux/fig_pc_case}

Overall, we make the following contributions:
\begin{enumerate}
  \item We propose a performance-anomaly-driven autoresearch flow that organizes
  specialized agents to localize residual misses, diagnose their causes, construct
  runnable minimal cases, and synthesize new prefetching mechanisms.
  \item We introduce \alectofull, a 17-engine prefetcher that uses residual gating,
  shared de-duplication, and accuracy throttling to integrate agent-generated
  specialists without disrupting effective existing mechanisms.
  \item Our evaluation shows that \alectofinal achieves a 61.1\% geomean IPC speedup
  over no prefetching on held-out workloads, outperforming Alecto, Berti, and Pythia
  by 14.5\%, 21.6\%, and 23.6\%, respectively, while remaining RTL-practical.
  \item To our knowledge, this is the first empirical demonstration that an
  agent-driven hardware-design process can produce an RTL-practical prefetcher that
  outperforms state-of-the-art human designs on unseen workloads.
\end{enumerate}

%% file: src/aux/fig_pc_case.tex
\begin{figure*}[!t]
  \centering
  \includegraphics[width=7in]{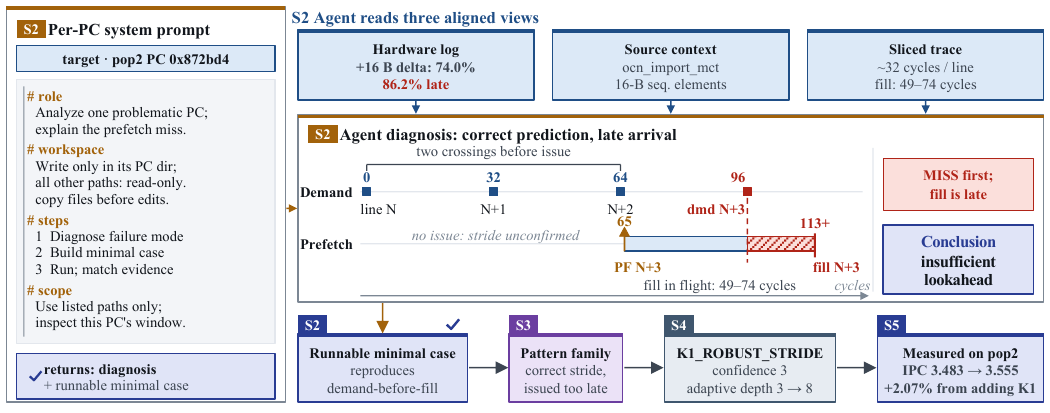}
  \caption{A complete S2--S5 pass for \texttt{pop2} PC \texttt{0x872bd4}. The S2
  prompt confines the agent to one PC and requires a diagnosis plus a runnable
  minimal case. Three aligned views show a dominant $+16$-byte stride (74.0\%),
  sequential 16-byte elements, and roughly 32 cycles between line crossings, yet
  86.2\% of prefetches are late because a 49--74-cycle fill starts only after two
  crossings confirm the stride. The timeline exposes the mismatch: demand reaches
  line $N{+}3$ at cycle 96, but its prefetch cannot fill before cycle 113. The
  reproducer therefore validates correct prediction with insufficient lookahead,
  not an accuracy failure. S3 classifies this late-stride family; S4 synthesizes
  \texttt{K1\_ROBUST\_STRIDE}, retaining confidence three while adapting depth from
  three toward eight; and S5 accepts it after \texttt{pop2} IPC rises from 3.483 to
  3.555 (2.07\%).}
  \label{fig:pc-case}
\end{figure*}

%% file: src/background_llm.tex
\section{Background}
\label{sec:bg}

\alectofinal combines agent-driven hardware design with ensemble prefetching. We next
review agent-based generate--evaluate loops, specialized prefetching mechanisms, and
their integration in ensembles.

\subsection{AI Agent-Driven Architecture Design}
\label{sec:bg-llm}

Our construction methodology builds on a body of work that uses large language models
and automated search to take over human design effort, from placing macros
competitively with experts~\cite{mirhoseini2021chip} and assisting
register-transfer-level coding and electronic design automation~\cite{liu2023chipnemo}
to grading machine-generated Verilog~\cite{liu2023verilogeval,thakur2023verilog} and
exploring accelerator design spaces from natural-language
specifications~\cite{fu2023gpt4aigchip}. The lesson common to these results is that the
generator need not be correct on its own. Given a faithful evaluator that scores each
candidate, even an imperfect model reaches expert quality by proposing and being
filtered. Prefetching fits this paradigm unusually well, since a cycle-accurate trace
simulator is exactly such an evaluator and the residual misses a candidate leaves behind
give a precise and localizable fitness signal.

A complementary line places language models inside an evolutionary loop to
\emph{discover} algorithms rather than transcribe them, producing new constructions for
open mathematical problems~\cite{romera2024funsearch}, sorting routines that shipped in
standard libraries~\cite{mankowitz2023alphadev}, programs evolved against automated
fitness functions~\cite{novikov2025alphaevolve,openevolve2025}, and reward functions for
downstream optimizers~\cite{ma2024eureka}. We adopt this loop, yet our contribution is
neither a new model nor a new search operator. It is the measurement-driven flow of
Section~\ref{sec:flow} that aims the loop at the anomalies a deployed prefetcher leaves
uncovered, and the non-interfering substrate of Section~\ref{sec:scaling} that lets
generated engines accumulate into a prefetcher scaling with the number of integrated
engines rather than into a pile of conflicting heuristics.

%% file: src/background_prefetching.tex
\subsection{Hardware Prefetching}
\label{sec:bg-prefetch}

A hardware prefetcher predicts future references and fetches the corresponding lines
ahead of demand; its quality is measured by accuracy, coverage, and timeliness.
Decades of work give a catalog of mechanism families, each keyed to one structural
regularity: per-instruction stride and sequential-stream
detectors~\cite{chen1995effective,jouppi1990stream,smith1982cache}, with best-offset
and SPP variants~\cite{michaud2016bop,kim2016spp} and the local-delta
Berti~\cite{navarro2022berti}; spatial footprint
replay~\cite{somogyi2006sms,ishii2009ampm,bakhshalipour2019bingo}; temporal and
delta correlation~\cite{nesbit2004ghb,joseph1997markov,jain2013isb,
bakhshalipour2018domino,grannaes2011dcpt,shevgoor2015vldp,yu2015imp}; and learned
address or policy models~\cite{hashemi2018learning,shi2021voyager,bera2021pythia}.
Each is strong on its target pattern and weak off it.

\subsection{Prefetching Ensemble}
Real programs interleave regular, spatial, temporal, and irregular access across
phases, so a prefetcher specialized for one family inevitably leaves useful coverage
to others~\cite{ayers2020classifying}. Research prototypes and shipping processors
therefore combine several specialized prefetchers and let their coverage add:
division-of-labor designs split prediction across cooperating
engines~\cite{kondguli2018division}, instruction-pointer classifiers route each
access to a matched specialist~\cite{pakalapati2020bouquet}, and commercial cores
deploy multiple stride, stream, and spatial prefetchers per
core~\cite{intel2023optref}. Such ensembles routinely beat the best single
prefetcher, yet composition is hard: the engines contend for cache capacity, off-chip
bandwidth, and miss-status holding registers, and naive combination regresses
somewhere even when it helps on average. Run-time selection admits one safe candidate
at a time~\cite{pugsley2014sandbox}, but keeps a single winner and discards the
coverage of the rest. Missing is a discipline under which many engines issue
\emph{simultaneously} without interference. This work builds directly on the ensemble line and takes one such ensemble,
Alecto~\cite{li2025alecto}, as its starting point. Alecto is a strong L1 data
prefetcher that already composes three base engines, namely a spatial pattern-merging
engine, a confirmed constant-stride engine, and a global stride engine, behind an
allocation-table state machine that learns, for each instruction pointer, which engine
suits the current data flow and routes training and issue accordingly. We adopt a strong
ensemble rather than a weak baseline on purpose, since lifting an already strong starting
point is far more convincing evidence for the flow than lifting a weak one. The rest of
this paper grows new sub-prefetchers on top of Alecto. The scaling result
of Section~\ref{sec:scaling}, the anomaly-driven discovery flow of
Section~\ref{sec:flow}, and the new-pattern engines of Section~\ref{sec:newpattern} are
all added to it, and the fully composed result is \alectofinal from
Section~\ref{sec:scaling}.

%% file: src/flow.tex
\section{The Performance-Anomaly-Driven Evolve Flow}
\label{sec:flow}

\begin{figure*}[t]
  \centering
  \includegraphics[width=7in]{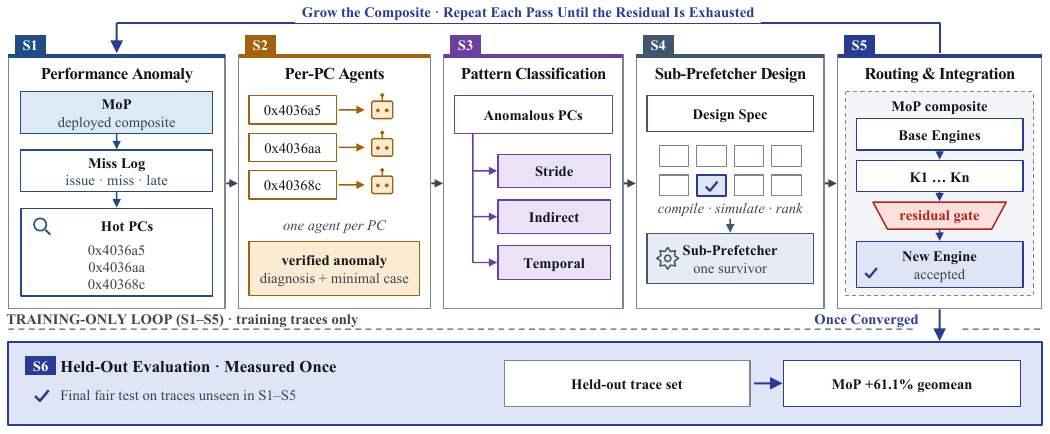}
  \caption{The performance-anomaly-driven autoresearch loop and its train/test
  boundary. The upper row is training-only: S1 reduces the deployed composite's
  residual misses to hot PCs; S2 admits only per-PC diagnoses backed by runnable
  minimal cases; S3 merges verified anomalies into pattern families; S4 compiles,
  simulates, and ranks candidate sub-prefetchers; and S5 inserts the survivor behind
  the residual gate. The upper feedback arrow carries this stronger composite---not
  held-out results---into the next S1 pass, so every iteration targets what the
  deployed design still fails to cover. Only after convergence does the downward
  arrow cross the dashed boundary into S6, which measures the frozen composite once
  on unseen held-out traces and reports its 61.1\% geomean IPC speedup over no
  prefetching. The figure thus separates iterative mechanism discovery from the
  single final generalization measurement.}
  \label{fig:flow}
\end{figure*}

The Performance-Anomaly-Driven Evolve Flow converts the residual misses of a deployed
prefetcher into a subprefetcher, as summarized in Figure~\ref{fig:flow}. It consists of six
stages. In S1, Performance Anomaly, we simulate the current prefetcher, record its miss
behavior, and identify the hot miss program counters. In S2, Per-PC Agent Analysis, we
invoke one agent per hot program counter to diagnose why it misses and to distill a minimal
reduced case. In S3, Pattern Classification, we group the analyzed program counters into
pattern families. In S4, Sub-Prefetcher Design, we synthesize a sub-prefetcher targeted at
each pattern family and test it in isolation to confirm that it covers that family. In S5,
Routing and Integration, we fold the new sub-prefetcher into the deployed composite,
tune its routing, and verify on the training traces that it covers only its intended pattern without disturbing
the other sub-prefetchers. In S6, Held-out Evaluation, we measure the composite on a
held-out trace set that took no part in the design. Throughout this section we thread one
real pass through all six stages as a running example, the pass that produced the
\texttt{K1\_ROBUST\_STRIDE} engine, so that the general loop becomes legible on a concrete
residual.

Figure~\ref{fig:pc-case} previews the compressed per-PC system prompt and follows
one program counter from its measured anomaly through the synthesized repair and
IPC outcome.

\noindent\textbf{S1: Performance Anomaly.}
S1 turns the deployed composite into a ranked list of the program counters where it
still misses. We simulate the deployed composite on the training traces and record, for every demand
access, whether it hit or missed, which engine if any covered it, and how timely that
prefetch was. Attributing each residual miss to the instruction pointer that produced
it is what makes the rest of the flow tractable, since we then spend effort only where
the composite is demonstrably failing rather than chasing an aggregate miss rate. The
event log is consumed and discarded incrementally, so the pass costs simulation time
but almost no storage. It ends with a miss log and a short list of hot miss PCs such
as \texttt{0x40249a} in 462.libquantum, the work items that S2 analyzes one at a time.

\noindent\textbf{Example.} On the pass that built \texttt{K1\_ROBUST\_STRIDE}, S1 runs
after the base engines and the broad streaming and region sub-prefetchers are already in
place, so the easy sequential coverage is done and only harder residuals remain. The
profiler does not ask which workload is slow but which demand misses survive that roster,
and it reduces the event stream to forty-four hot residual-miss program counters. At many
of these sites the composite is far from idle: it issues hundreds of thousands of
prefetches with correct addresses, yet a large fraction arrive after the demand has
already reached the line, and almost nothing is evicted unused. High issue volume with
correct addresses but late arrival is the first sign that the failure is timing rather
than accuracy, and these forty-four pointers are what the per-PC agents take up next.

\noindent\textbf{S2: Per-PC Agent Analysis.}
S2 diagnoses why each hot PC misses by handing it to its own analysis agent, one per
program counter. A residual miss has several causes an aggregate statistic cannot tell
apart: the composite may issue nothing for the PC, issue inaccurately, or issue too
late, and only the first calls for a new mechanism while the others call for a routing
change. Separating these symptoms needs both low-level hardware evidence and
high-level program meaning, so each agent reads three read-only inputs, namely the
hardware log for its PC, the source code, and a sliced trace, and is confined to a
per-PC workspace so its reasoning never leaks across pointers.
Figure~\ref{fig:pc-case} shows this contract for the real \texttt{pop2} PC
\texttt{0x872bd4}: the agent must return an evidence-backed diagnosis with a runnable
minimal reproducer before any finding is accepted. The reproducer forces each
diagnosis to be checked against the event evidence rather than asserted, and it is
this reduced case that S3 classifies.

\noindent\textbf{Example.} Figure~\ref{fig:pc-case} follows the pop2 member of this
family through the three aligned evidence views, the late-arrival diagnosis, its runnable
minimal case, and the resulting \texttt{K1\_ROBUST\_STRIDE} repair. On this pass, one
agent per hot program counter reconstructs why its correct predictions land late, and
three program counters from the training workloads tell the recurring story. At
libquantum \texttt{0x401442} the stride engine tracks a clean
64-byte line delta at 62.5\% dominance and issues 910{,}546 prefetches, yet its
degree-three lookahead reaches only three lines, 192 bytes, ahead of the demand, far too
few to hide the roughly 120-cycle fill when a burst takes six demand misses in a single
cycle, so 69.9\% of those prefetches arrive after the line is needed, and only twelve are
ever evicted unused. At pop2 \texttt{0x872bd4} the engine detects a +16-byte element
stride at 74.0\% dominance but withholds issue until two consecutive line crossings
confirm it, by which point the demand has already advanced two lines, and 86.2\% of the
prefetched lines land after the access that wanted them. At libquantum \texttt{0x4011ae}
the 64-byte stride is confirmed at 77.1\% dominance and the engine issues 497{,}981
prefetches, but the core runs its reorder buffer 97\% full with its miss-status registers
saturated and drains queued loads faster than the fills return, so 71.4\% arrive late.
Each agent confirms its diagnosis with a minimal reproducer whose late fraction matches
the hardware within a few points, so the verdict is evidence-backed rather than asserted:
the engines are correct but shallow, and the three sites share one shape.

\noindent\textbf{S3: Pattern Classification.}
S3 clusters the diagnosed program counters into a few pattern families so that a
single engine can cover an entire class rather than one PC at a time. For each anomaly
we read the evidence S2 produced, its line-delta histogram, its delta dominance, the
number of streams aliased under it, and its occupancy of the reorder buffer, MSHRs,
and prefetch queue, and assign it to a family: a confirmed stride that arrives late,
one pointer aliasing several interleaved streams, a short fixed-delta fan-out behind a
large jump, or an irregular temporal recurrence with no learnable successor. Collapsing
many program counters into stride, indirect, and temporal classes is what keeps the
engine count bounded as the loop runs.
Anomalies with no learnable structure are not forced into a family; we set them aside
for the discovery analysis of Section~\ref{sec:newpattern}, while the rest pass to S4
as design targets.

\noindent\textbf{Example.} The three diagnosed pointers cluster cleanly. From each
anomaly's line-delta histogram and dominance, its aliased-stream count, and its
reorder-buffer, MSHR, and prefetch-queue occupancy, the stride-confirmed but late program
counters across libquantum and pop2 collapse into a single family, insufficient lookahead
with prefetch lateness. The label is discriminative, not nominal. These sites are not the
unpredictable family, since each is driven by a dominant arithmetic stride; not the
capacity-defeated family, since almost nothing is evicted unused; not the multi-stream
family, since a single stride dominates rather than the interleaved page-walk and data
streams that poison a neighbor such as leslie3d \texttt{0x40d842}; and not the
translation-dominated family, since the misses are genuine data misses. What unites them,
and only them, is a confirmed stride issued too late. That label selects the response: the
family needs no new access pattern, only the confirmed stride run further ahead while the
stream bursts, with the depth retracted once the prediction stops paying off.

\noindent\textbf{S4: Sub-Prefetcher Design.}
S4 synthesizes one sub-prefetcher per family by searching for an implementation rather
than writing it by hand. Each family yields a design specification fixing the trigger
condition, the prediction rule, and the integration interface, which we hand to an
evolutionary code-generation loop~\cite{openevolve2025} that proposes candidates,
compiles them against the simulator, and ranks them by a fitness function defined on
training traces only. Searching lets the loop discover parameters such as the
adaptive run-ahead depth behind 462.libquantum's late stride that are tedious to tune
by hand, while the training-only fitness leaves the held-out set untouched. The loop
reuses prior cases,
so a family an existing engine already covers is not re-synthesized, and it unit-tests
each candidate in isolation to confirm coverage. The surviving candidate enters S5 as
a single new engine.

\noindent\textbf{Example.} For this family the specification becomes one engine,
\texttt{K1\_ROBUST\_STRIDE}. It keeps a per-program-counter stride table and promotes an
entry only after the same line delta repeats past a confidence threshold of three. Once
the stride is confirmed it issues an adaptive run-ahead that escalates from a depth of
three lines toward eight as it observes its own prefetches arriving late, so the lookahead
grows to cover the burst rather than the first few lines. It trains on every demand access
so it observes the true stride, issues only for accesses the base engines left uncovered,
and a per-source accuracy throttle disables it if its confirmed fraction collapses, which
stops the deeper run-ahead from wasting bandwidth where the stride is not real.
Unit-tested in isolation on the training stream, the single engine turns the late issues
at all three member pointers into timely ones and the family's residual clusters
disappear, confirming that the classification captured one mechanism and not a coincidence
of unrelated sites.

\noindent\textbf{S5: Routing and Integration.}
S5 folds the surviving sub-prefetcher back into the deployed composite and verifies,
on training traces only, that it repairs its target pattern without disturbing the engines already in
place. We add it under the residual-gating discipline of Section~\ref{sec:scaling}, so
it is consulted only on accesses the composite still misses, and we tune its routing
until it fires on its own family and stays silent elsewhere. Confirming
non-interference before any held-out measurement is what lets each pass add a strictly
non-interfering step to the composite. This closes the training-only loop, and S1 through
S5 repeat
until the residual is exhausted.

\noindent\textbf{Example.} Folding \texttt{K1\_ROBUST\_STRIDE} into the composite, the
repair lands on exactly the family members it was built for: the late issues at the
libquantum and pop2 sites become timely, and no other training workload moves. The engine
thus adds one non-interfering riser to the staircase of Section~\ref{sec:scaling}, the
signal that it causes no regression before any held-out trace is touched.

\noindent\textbf{S6: Held-out Evaluation.}
S6 is the only stage that touches held-out traces, and it exists to keep every reported
speedup honest. A recurring hazard in measurement-driven design is that the traces used
to discover a mechanism also appear in its evaluation, which inflates the reported
benefit. Our flow keeps a single source of truth for the train and held-out partition,
and stages S1 through S5 assert at run time that no held-out trace is present, so a leak
fails loudly. Once the loop has converged, S6 runs the final composite on traces it
never observed, and \alectofinal{}'s $+61.1\%$ geomean speedup of Section~\ref{sec:eval} is
measured here and nowhere else.

\noindent\textbf{Example.} Only now do we measure \texttt{K1\_ROBUST\_STRIDE} on the
held-out trace set that took no part in the design. It lifts the held-out geomean speedup
over the base by 1.10 points and takes no workload backward, and the gains land on codes
the engine never observed, with fotonik3d up 8.66 points, bwaves up 0.93, lbm up 0.78, and
cam4 up 0.66, so the mechanism diagnosed on the libquantum and pop2 sites generalizes well
beyond them. In one pass the flow localized a residual that looked like a solved pattern,
clustered three independent late-stride sites across two training workloads into one
family, recognized that the shared failure was timing rather than coverage, synthesized a
single engine that deepens its reach only when its predictions are confirmed, and verified
on unseen traces that the fix holds.

%% file: src/scaling.tex
\section{Three Challenges for a Self-Improving Prefetcher}
\label{sec:scaling}

Three challenges separate an agent loop that merely searches from one that
produces a prefetcher that actually improves. \textbf{Composition must be
monotone.} On a strong baseline, letting an agent edit the prefetcher freely
does not scale. When the agent is allowed large edits, it tends to rewrite the
strong baseline destructively, so accesses that the baseline used to cover
become uncovered and improvements never accumulate. When the agent is restricted
to small edits, it only retunes existing parameters and the headroom is small.
Escaping this dilemma requires a substrate that makes integration monotone by
construction, which we develop in Section~\ref{sec:scaling-c1}. \textbf{Search
must be anomaly-driven.} A loop that searches a fixed textbook library is a
glorified ensemble. Ours is performance-anomaly-driven. It spends effort where
the deployed prefetcher is anomalously ineffective, and when the residual
resists every library mechanism it synthesizes a genuinely new engine instead of
recombining old ones, as we show in Section~\ref{sec:newpattern}. \textbf{Diagnosis
must be reducible.} An agent cannot explain, let alone repair, a residual miss while
it is buried in a workload of billions of instructions and tens of millions of miss
events, so each anomaly must be distilled to a minimal case that re-exhibits the same
failure and can be optimized in isolation, as we develop in
Section~\ref{sec:scaling-c3}.

\subsection{For Challenge 1: Scale-Out Incremental Mechanism}
\label{sec:scaling-c1}

The flow must guarantee that successive iterations never damage a mechanism the
agent has already built, yet still add enough coverage to move the needle. Our
key insight is that the base prefetcher must be loosely coupled, so that each
iteration contributes a self-contained sub-module and a routing layer decides
where every module trains and issues. We argue that an ensemble prefetcher is
the natural substrate for this discipline. Each iteration appends one
sub-prefetcher to the existing roster, and the router folds it in, so no
iteration rewrites a working engine while new mechanisms remain free to cover new
patterns. This is what lets the composite scale out one engine at a time.

We take Alecto as the starting point for two reasons. First, it is a
strong baseline, and lifting a strong starting point is far more convincing
evidence for the flow than lifting a weak one. Second, it is already an
ensemble. As Table~\ref{tab:roster} shows, it composes three base engines, namely
a spatial pattern-merging engine, a confirmed constant-stride engine, and a
global stride engine, together with an allocation-table state machine that learns,
for each instruction pointer, which engine suits the current data flow and routes
training and issue accordingly.

However, every sub-prefetcher is an extra engine layered on the base, and left
unchecked it pollutes the cache and saturates the prefetch queue and the MSHRs,
erasing the benefit of engines already working. Three mechanisms prevent this.
\emph{Residual gating:} every sub-prefetcher trains on every access but may
\emph{issue} only when the base and all earlier sub-prefetchers issued nothing
for that access, a condition we call uncovered, so a new engine fills gaps and
never displaces a working one, which is the structural reason the composite is
monotone. \emph{Cross-engine sandbox de-duplication:} a shared sandbox records
recently requested lines and suppresses duplicates before they reach the cache,
so an engine's marginal cost tracks the unique coverage it adds rather than its
raw issue rate. \emph{Ground-truth accuracy throttle:} a windowed accuracy
computed from the cache's ground-truth useful and fill signals disables the
extension layers and falls back to the base whenever that accuracy collapses.

\begin{figure*}[t]
  \centering
  \includegraphics[width=0.92\textwidth]{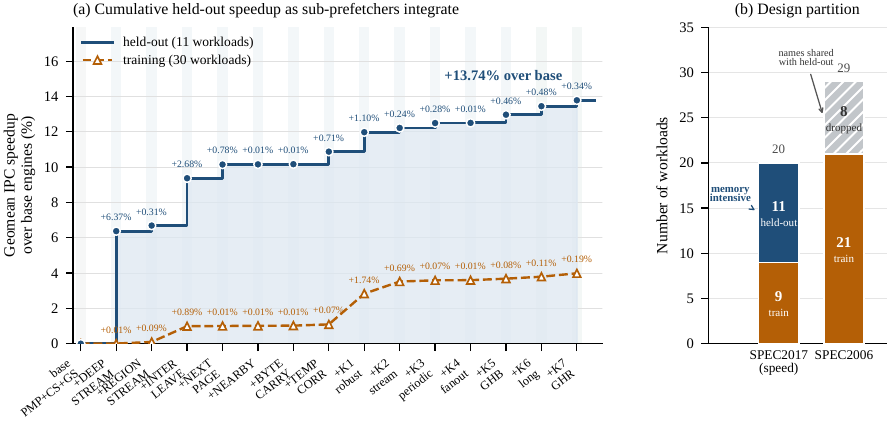}
  \caption{\alectofinal scales by composition. Panel (a) integrates the sub-prefetchers
  one engine at a time and plots the cumulative geomean IPC speedup over the base
  engines, with the held-out curve in blue and the training curve in orange. Both
  staircases rise monotonically and neither regresses at any step, because residual
  gating keeps each addition non-interfering. The held-out curve reaches 13.74\% over
  the base and the training curve 3.98\%. The held-out gain is more than three times the
  training gain, so the composition does not merely transfer to unseen workloads but
  improves most on the distribution it never saw, which we read as out-of-distribution
  generalization and as direct evidence for the generality of the discovery flow rather
  than of any single benchmark. Panel (b) shows
  the design partition: the eleven memory-intensive SPECspeed2017 traces are held out,
  while the nine remaining SPECspeed2017 traces and twenty-one SPEC CPU2006 traces form
  the thirty-workload training set, and the eight SPEC CPU2006 benchmarks whose names
  recur in SPECspeed2017 are dropped so that no held-out identity leaks into design.}
  \label{fig:scaling}
\end{figure*}

\noindent\textbf{Result.} Figure~\ref{fig:scaling} reports this paradigm. We
integrate the sub-prefetchers one engine at a time. On the training traces the
orange curve rises monotonically and never regresses, and on held-out traces never
seen during training the blue curve does the same, reaching 13.74\% over the base
engines, which confirms that the flow does not overfit. The final two steps come
from the new-pattern engines that the discovery flow synthesizes, which we
present in Section~\ref{sec:newpattern}. Table~\ref{tab:roster} lists the fully
composed \alectofinal: three base engines with source identifiers one through
three and fourteen sub-prefetchers with identifiers four through seventeen, and each
sub-prefetcher covers a pattern the others miss and contributes its own
incremental gain.

\begin{table*}[t]
  \centering
  \caption{Sub-prefetchers in the fully composed \alectofinal. Source
  identifiers 1--3 are base engines and 4--17 are
  sub-prefetchers.}
  \label{tab:roster}
  \resizebox{0.96\textwidth}{!}{\includegraphics[width=\textwidth]{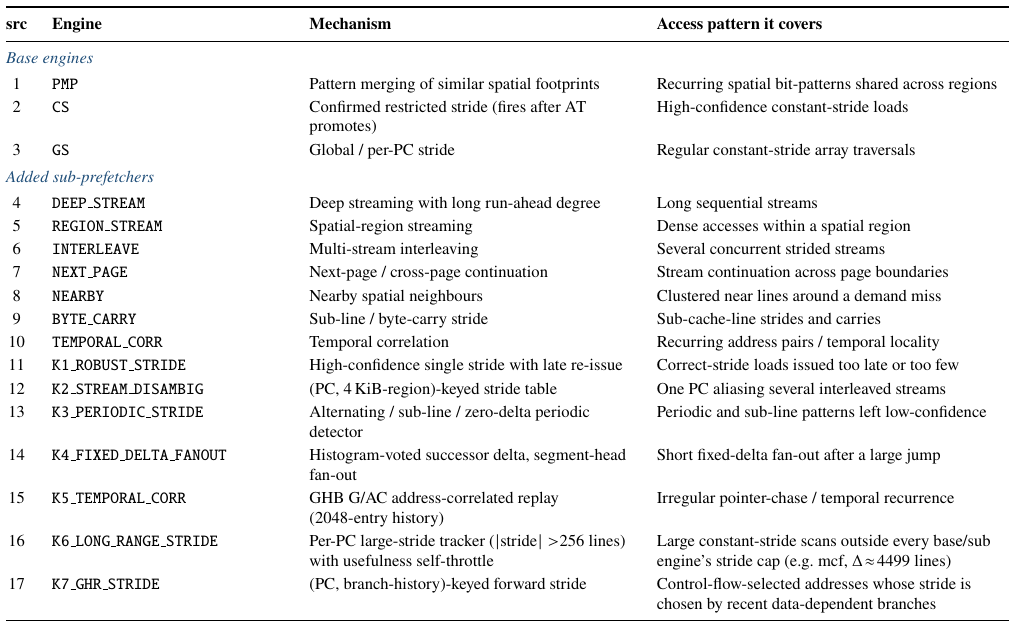}}
\end{table*}

\subsection{For Challenge 2: Performance-Anomaly-Driven Flow}
\label{sec:newpattern}

Monotone composition alone is not enough. If the agent simply iterates on this
substrate, it has no sense of direction and wanders the design space freely, and
it tends to recombine the existing sub-prefetchers by rote rather than surface a
pattern none of them captures. We therefore drive the loop with the performance
anomaly itself. Each iteration targets the patterns the current prefetcher leaves
uncovered, which gives the agent both a direction and the context it needs,
namely the hardware residual-miss log and the high-level source code that
together explain why those accesses missed. Crucially, this turns the loop into a
discovery instrument rather than a recombination engine. One example makes the
point.

The flow surfaces patterns that no program-counter history can predict.
Conditioning a per-instruction stride table on a gshare-style global
branch-history register exposes a pattern outside the entire address-history
family: loads whose forward stride is selected by recent data-dependent branches.
A per-instruction single stride explains only 30 to 48\% of these deltas,
whereas the same table indexed by six to twelve bits of branch history explains
76 to 92\%. A random key of equal cardinality stays at the single-stride
baseline, so the branch history carries real information rather than merely
enlarging the table. This signal sits outside the mainstream feature set. Stride,
Berti, SPP, AMPM, SMS, temporal prefetchers, and even the learned Pythia policy
all key on program counter, delta, and page offset rather than on branch
history~\cite{navarro2022berti,kim2016spp,ishii2009ampm,somogyi2006sms,bera2021pythia}.
Branch-directed prefetching has been explored through runahead and register
tracking~\cite{kim2014bfetch}, yet a lightweight branch-history-indexed stride
table is a simpler instantiation and, to our knowledge, the first surfaced
automatically from a deployed prefetcher's residual. The engine trains on every
access, issues only on base-uncovered misses, and suppresses any prediction that
matches the program counter's stable dominant stride. Integrated as the
fourteenth sub-prefetcher, it contributes \textbf{0.34\%}, concentrated where
control flow steers data addresses. The two most branch-correlated workloads,
649.fotonik3d and 602.gcc, gain 5.0\% and 1.0\% respectively while the rest stay
flat. Its predictive variable was absent from every existing engine, so pure
composition could never have reached it.

\subsection{For Challenge 3: Minimum Case Reduction}
\label{sec:scaling-c3}

A directed search is only as trustworthy as the diagnoses it acts on, and a residual
miss does not arrive as a clean question. It is buried in a workload of billions of
instructions and tens of millions of miss events, entangled with every other access
the program makes, so an agent asked to explain it in place can assert any plausible
story and nothing forces that story to be correct. Our key insight is that every
diagnosis must be reduced to a minimal case that re-exhibits the same failure on its
own, which turns the explanation from a claim into a runnable artifact. S2 of the flow
in Section~\ref{sec:flow} therefore requires each per-PC agent to distill its anomaly
into a minimal case that is analyzable, reproducible, and optimizable in isolation.

Three properties make the reduced case load-bearing. First, it is \emph{analyzable}: the
agent works in the per-PC workspace of Figure~\ref{fig:pc-case}, reads only the hardware
log for its program counter, the source line, and a sliced trace, and writes a reproducer
of a few hundred lines that a person can read end to end rather than reasoning over the
full event stream. Second, it is \emph{reproducible}: the reproducer is self-contained, a
seeded synthetic generator or a byte-exact trace window shipped with the exact commands to
rerun it, so the failure reappears deterministically instead of as a one-off measurement.
Third, it is \emph{optimizable}: because the case is isolated, a candidate fix can be
tested on it before it ever touches the composite, which is precisely the in-isolation
unit test S4 runs on each synthesized engine.

\noindent\textbf{Example.} Consider bzip2 program counter \texttt{0x404b36}, the inner
comparison load of a three-way quicksort partition. The anomaly arrives as a
38-million-instruction window of a 7-billion-instruction trace paired with a
50-million-line event log, far more than an agent can reason over directly. The agent
reduces it to a 159-line seeded trace generator and a 239-line cache model that together
reproduce the access pattern. Re-run through the same simulator, the synthetic case
collapses to a 3.5\% L1D hit rate and lets only 10.5\% of its prefetch requests reach the
cache, the same queue saturation recorded as 3{,}013 prefetch-queue-full events in the
original. That match falsifies the obvious missing-pattern story and pins the failure on
cache capacity and queue pressure, a routing and throttling problem rather than a new
access pattern, so the flow spends no engine where none would help.

Without this step the per-PC analyses would be a pile of unfalsifiable assertions, and a
flow that builds engines from unverified diagnoses would amplify its own mistakes one
iteration at a time. The minimal case is the checkpoint that forces every downstream
stage, the classification of S3, the design of S4, and the integration of S5, to rest on
a cause that has been reproduced rather than a story that merely sounds right.

%% file: src/methodology.tex
\section{Experiment}
\label{sec:method}

\subsection{ChampSim Setup}
\label{sec:method-champsim}

We evaluate in ChampSim, the trace-driven simulator used by the data
prefetching championships~\cite{gober2022championship}. \alectofinal and all baseline
prefetchers are implemented as L1D prefetchers in ChampSim. Table~\ref{tab:experiment_cfg}
summarizes the experimental configuration. We choose this core configuration to align
the simulated processor with an industrial Zen 5-class high-performance core.

\input{src/aux/experiment_cfg}

\subsection{Workload and Baseline}
\label{sec:method-workload}

We draw traces from the SPEC CPU2006 and SPEC CPU2017 benchmark suites and maintain a
single source of truth for the partition into training and held-out sets. The
held-out set comprises the eleven memory-intensive traces of SPECspeed2017 that never
participate in design, and the training set comprises thirty traces, namely twenty-one
traces from SPEC CPU2006 and the nine remaining compute-bound traces of SPECspeed2017.
The two SPECspeed2017 subsets exhaust the suite, and no benchmark name appears in both
the training and the held-out set: we drop the SPEC CPU2006 version of any benchmark
whose name recurs in SPECspeed2017 so that no held-out identity leaks into design. The
split therefore also stresses cross-suite generalization, since design never sees a
SPEC CPU2017 memory-intensive trace. Every reported speedup is measured on the held-out
set. Each simulation warms the caches for ten million instructions and then measures
one hundred million instructions.

The training workloads are bzip2, gamess, milc, zeusmp, gromacs, cactusADM,
leslie3d, namd, gobmk, dealII, soplex, povray, calculix, hmmer, sjeng,
GemsFDTD, libquantum, h264ref, tonto, astar, sphinx3, perlbench, wrf, x264,
pop2, deepsjeng, imagick, leela, nab, and exchange2. The held-out workloads are
gcc, bwaves, mcf, cactuBSSN, lbm, omnetpp, xalancbmk, cam4, fotonik3d, roms, and
xz.

We compare against no prefetching and against representative state-of-the-art single
prefetchers, namely SPP~\cite{kim2016spp},
Berti~\cite{navarro2022berti}, and Pythia~\cite{bera2021pythia}, which we evaluate in
the same framework under the identical held-out protocol and with the same-sized
prefetch resources. For a storage-fair comparison, we evaluate \alectobaseline with
its tables enlarged to the same 110\,KB budget as \alectofinal, isolating the
contribution of the residual sub-prefetcher layer from added storage capacity.

%% file: src/aux/experiment_cfg.tex
\begin{table}[t]
  \small
  \centering
  \caption{ChampSim experimental configuration. All prefetchers are evaluated in
  the same simulator with the same core, cache, memory, and prefetch-resource
  settings.}
  \begin{tabular}{@{}p{0.25\columnwidth}|p{0.65\columnwidth}@{}}
    \toprule
    \textbf{Module} & \textbf{Configuration} \\
    \midrule
    Simulator & ChampSim trace-driven simulation \newline
    One simulated core, 64B cache line \newline
    L1D prefetcher interface for all designs \\ \hline

    Core & 4GHz, 8-wide fetch/decode/issue/commit \newline
    448-entry ROB, 192/96-entry LQ/SQ \newline
    240-entry physical register file \newline
    192-entry scheduler \\ \hline

    Cache hierarchy & L1I: 32KB, 8-way, 4-cycle latency \newline
    L1D: 48KB, 12-way, 4-cycle latency \newline
    L2: 1MB, 16-way, 10-cycle latency \newline
    LLC: 4MB, 16-way, 20-cycle latency \\ \hline

    Memory & Two channels, one rank per channel \newline
    5600 MT/s DDR5 memory \\ \hline

    Prefetch resources & L1D prefetch queue: 64 entries \newline
    L1D MSHRs: 64 entries \newline
    L2 prefetch queue: 32 entries \newline
    L2 MSHRs: 48 entries \\ \hline

    Workloads & Training: 30 SPEC CPU2006 and CPU2017 traces \newline
    Held-out: 11 SPEC CPU2017 memory-intensive traces \newline
    10M warmup instructions + 100M measured instructions \\ \hline
    \bottomrule
  \end{tabular}
  \label{tab:experiment_cfg}
\end{table}

%% file: src/evaluation.tex
\section{Evaluation}
\label{sec:eval}

This section substantiates four claims. First, the composite scales, so that every
integrated engine improves the held-out geomean and none regresses it. Second, the fully
composed \alectofinal outperforms state-of-the-art single prefetchers, and it does so
across heterogeneous workloads rather than on a single benchmark. Third, this advantage
traces to demand-miss coverage of the union of pattern families rather than to dominance
of any one family. Finally, the design is realizable, carrying a modest storage and area
budget together with a one-time discovery cost that amortizes over the deployed datapath.
We treat each claim in turn.

\subsection{The Evolve Steps}
\label{sec:eval-scale}

Figure~\ref{fig:scaling} reads the composition off a single figure. We integrate the
sub-prefetchers one engine at a time, and the figure plots the cumulative geomean IPC
speedup over the base engines as each engine joins, with the held-out curve in blue and
the training curve in orange. Both curves are staircases that only step up. No engine
regresses either curve at any point, which is the direct empirical signature of residual
gating, since an engine that may issue only on uncovered accesses can fill a gap but
cannot displace an engine already working. Reading the endpoints, the held-out curve
reaches 13.74\% over the base while the training curve reaches 3.98\% over the base.
Because the base engines already deliver 41.6\% over no prefetching, this composition
lifts the fully composed \alectofinal to 61.1\% over no prefetching.

The two staircases together make a generalization claim stronger than transfer alone. The
held-out workloads are out of distribution by construction, as the design partition in
Figure~\ref{fig:scaling}b makes explicit. They are the memory-intensive traces of
SPECspeed2017 and the discovery flow never observes them, while design sees only SPEC
CPU2006 and the compute-bound remainder of SPECspeed2017, and we drop every SPEC CPU2006
benchmark whose name recurs in SPECspeed2017 so that no held-out identity can leak into the
loop. Crucially, the gain does not merely survive this distribution shift but grows under
it. The held-out geomean reaches 13.74\% against 3.98\% on the training set, more than three
times the training gain, so the flow helps the unseen distribution more than the one
it was built on. It is the signature of superior generalization in its strongest form.

\begin{figure*}[t]
  \centering
  \includegraphics[width=\textwidth]{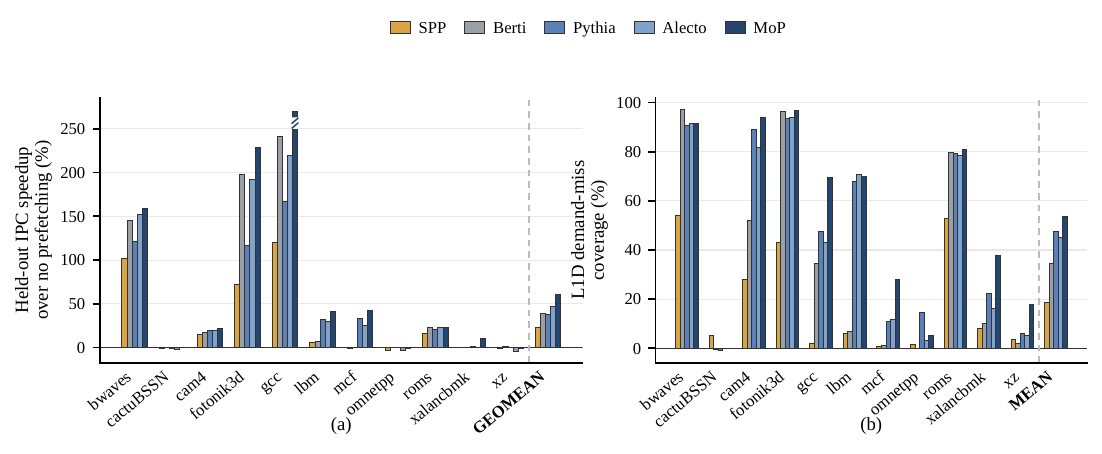}
  \caption{Held-out performance and miss coverage. The left panel (a)
  reports per-workload IPC speedup over no prefetching for \alectofinal,
  \alectobaseline, SPP, Pythia and Berti, where \alectobaseline uses the same
  110\,KB storage budget as \alectofinal. \alectofinal attains the best geomean and is never
  the worst on any workload, outperforming Berti by 21.6\%, Pythia by 23.6\%,
  and \alectobaseline by 14.5\%. The right panel (b) reports per-workload
  L1D demand-miss coverage.  \alectofinal matches the best baseline
  on the regular workloads and is the only design that also covers the irregular ones,
  namely mcf, omnetpp, and xalancbmk, giving the highest mean coverage.}
  \label{fig:sota}
\end{figure*}

\subsection{End to End Performance Evaluation}
\label{sec:eval-sota}

Figure~\ref{fig:sota} compares \alectofinal against three state-of-the-art
prefetchers, SPP~\cite{kim2016spp},
Berti~\cite{navarro2022berti}, and Pythia~\cite{bera2021pythia}, under the identical
held-out protocol and with the same-sized prefetch resources for every design. We also
include the storage-equalized \alectobaseline, configured with the same 110\,KB budget
as \alectofinal, so that the evaluation separates added storage from the contribution
of our scale-out incremental mechanism. \alectofinal reaches a held-out
geomean speedup of \textbf{61.1\%} over no prefetching. This final design improves
over Berti by 21.6\%, over Pythia by 23.6\%, over
\alectobaseline by 14.5\%.

Two conclusions follow from the figure. First, every single prefetcher fails on some pattern
family. Berti and SPP capture the streaming and stride
workloads such as bwaves and fotonik3d yet collapse to the no-prefetching baseline on
the irregular workloads mcf, omnetpp, and xalancbmk. Pythia adapts more broadly but
still trails on the streaming workloads where a dedicated stride engine excels. No
baseline is simultaneously strong on both regular and irregular access. Second,
\alectofinal is the only design that tracks the best baseline on the regular workloads and
also wins the irregular workloads, for example by more than 30\% on mcf where
the stride and stream baselines contribute essentially nothing. As a result \alectofinal is
never the worst prefetcher on any workload, and it attains the best geomean by
covering the union of the patterns that the individual baselines cover. This
per-workload robustness is exactly the behavior that scaling by composition is
designed to produce.


\subsection{Coverage}
\label{sec:eval-cat}

Figure~\ref{fig:sota} traces the per-workload speedup to its proximate cause in its
coverage panel, which reports L1D demand-miss coverage. On the regular workloads bwaves,
cam4, fotonik3d, and roms, \alectofinal matches the strongest baseline, and on the
irregular workloads where the stride and stream baselines eliminate almost no misses,
namely mcf, omnetpp, and xalancbmk, \alectofinal is the only design with non-trivial
coverage. The panel also includes \alectobaseline with the same 110\,KB storage budget
as \alectofinal. Averaged
over the held-out set \alectofinal removes \textbf{54\%} of demand misses, ahead of
Pythia at 47\%, \alectobaseline at 45\%, Berti at 35\%, and SPP at 19\%. This is the coverage analog of
the per-workload IPC result, since \alectofinal wins not by
dominating any one pattern but by covering the union.


\subsection{Storage and Area Overhead}
\label{sec:eval-storage}

A meta-prefetcher composing seventeen engines invites the concern that it buys
performance with an unreasonable storage budget. It does not.
Table~\ref{tab:storage} accounts the on-chip state of the composed \alectofinal at
hardware-realizable widths, namely sixteen-bit partial tags, thirty-two-bit truncated
addresses, twelve-bit strides, small counters, and 256-entry four-way predictor
tables, for a total of \textbf{about 110\,KB} of SRAM. Synthesizing a
register-transfer-level model of the residual layer with Synopsys Design Compiler in
the 6nm library at the typical corner, 0.75\,V, and 25$^\circ$C maps the tables to
eleven 256$\times$4$\times$80-bit OpenRAM macros of 0.0341\,mm$^2$, while
the synthesized control logic of the \emph{entire} layer, comprising all seventeen
engines, the per-PC routing table, the sandbox de-duplication, and the residual arbiter,
is only 0.000565\,mm$^2$, so the composite occupies 0.0347\,mm$^2$ and is just 1.6\%
logic.

\begin{table}[t]
  \centering
  \caption{On-chip storage and 6nm area of the fully composed \alectofinal,
  synthesized with Synopsys Design Compiler at the typical corner at 0.75\,V and
  25$^\circ$C. \textbf{SRAM} is the N6 OpenRAM macro area apportioned to each engine's
  table budget, and \textbf{Logic} is the synthesized standard-cell area of each
  engine module reported hierarchically.}
  \label{tab:storage}
  \scriptsize
  \begin{tabular}{@{}l r r r@{}}
    \toprule
    \textbf{Component} & \textbf{Storage} & \textbf{SRAM} & \textbf{Logic} \\
                       & \textbf{(KB)}    & \textbf{(mm$^2$)} & \textbf{(mm$^2$)} \\
    \midrule
    Shared routing / allocation / sample table & 4.0 & 0.0012 & 0.0000135 \\
    PMP (src 1)                                & 1.0 & 0.0003 & 0.0000396 \\
    CS (src 2)                                 & 0.7 & 0.0002 & 0.0000436 \\
    GS / RST (src 3)                           & 1.3 & 0.0004 & 0.0000140 \\
    DEEP\_STREAM (src 4)                       & 2.4 & 0.0007 & 0.0000102 \\
    REGION\_STREAM (src 5)                     & 2.0 & 0.0006 & 0.0000100 \\
    INTERLEAVE (src 6)                         & 2.4 & 0.0007 & 0.0000156 \\
    NEXT\_PAGE (src 7)                         & 1.8 & 0.0006 & 0.0000117 \\
    NEARBY (src 8)                             & 1.4 & 0.0004 & 0.0000130 \\
    BYTE\_CARRY (src 9)                        & 4.8 & 0.0015 & 0.0000345 \\
    TEMPORAL\_CORR (src 10)                    & 12.0 & 0.0037 & 0.0000107 \\
    K1\_ROBUST\_STRIDE (src 11)                & 3.0 & 0.0009 & 0.0000469 \\
    K2\_STREAM\_DISAMBIG (src 12)              & 5.2 & 0.0016 & 0.0000418 \\
    K3\_PERIODIC\_STRIDE (src 13)              & 4.8 & 0.0015 & 0.0000424 \\
    K4\_FIXED\_DELTA\_FANOUT (src 14)          & 7.0 & 0.0022 & 0.0000366 \\
    K5\_TEMPORAL\_CORR (src 15)                & 19.0 & 0.0059 & 0.0000108 \\
    K6\_LONG\_RANGE\_STRIDE (src 16)           & 4.2 & 0.0013 & 0.0000637 \\
    K7\_GHR\_STRIDE (src 17)                   & 28.0 & 0.0087 & 0.0000445 \\
    Shared sandbox (de-duplication) + GHR      & 5.0 & 0.0015 & 0.0000057 \\
    Residual arbiter + top-level glue          & --- & ---    & 0.0000562 \\
    \midrule
    \textbf{Total}                             & \textbf{110.0} & \textbf{0.0341} & \textbf{0.0005650} \\
    \bottomrule
  \end{tabular}
\end{table}

Three points follow. The budget is dominated by two intrinsically table-heavy
engines, the K5 temporal correlation table at 19.0\,KB and the branch-history-indexed K7
table at 28\,KB, whose cost is inherent to the pointer-chasing and control-flow-selected
patterns they target, while every other engine together with all routing and
de-duplication state fits in roughly 63\,KB combined. The \emph{marginal} cost of an
added engine is small, a median per-engine table of two to five kilobytes, because all
engines share one routing table and one de-duplication sandbox, so engine count does not
multiply shared state, which is the storage analog of the non-interference result of
Section~\ref{sec:scaling-c1}. And the absolute budget is modest, since at 110\,KB the
composite is a few times the 25.5\,KB of a single learned policy such as
Pythia~\cite{bera2021pythia} yet subsumes an entire roster, and it stays one to two
orders of magnitude below dedicated temporal prefetchers whose metadata runs to hundreds
of kilobytes or megabytes~\cite{jain2013isb,wu2019triage}.

\subsection{Token Cost of the Discovery Flow}
\label{sec:eval-tokencost}

\begin{figure}[t]
  \centering
  \includegraphics[width=\columnwidth]{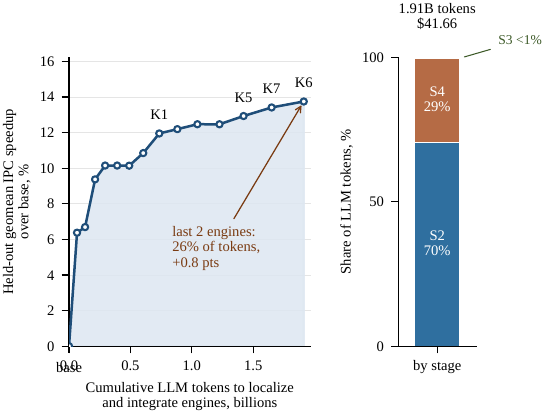}
  \caption{Discovery-flow token cost. The total and the metered \$41.66 are read from the
  DeepSeek V4 Pro usage export of the full campaign, which consumed 1.91 billion tokens;
  the two breakdowns are modeled and calibrated from completed runs. The per-PC analysis
  of S2 dominates the budget at 70\%, the synthesis of S4 adds 29\%, and the
  classification of S3 is below 1\%. The curve shows a clear diminishing return, since the
  early cheap engines deliver most of the speedup while the two new-pattern engines spend
  about a quarter of the tokens for the final 0.8 points.}
  \label{fig:token-cost}
\end{figure}

All agents in the flow run on DeepSeek V4 Pro, and Figure~\ref{fig:token-cost} reports its
token budget. Across the full campaign the flow consumed 1.91 billion tokens, billed at
\$41.66. Our analysis reveals that the cost is concentrated in a single stage. The per-PC
anomaly analysis of S2 accounts for about 70\% of the budget, the evolutionary synthesis
of S4 for a further 29\%, and the pattern classification of S3 is negligible at below
1\%, while the simulator-only stages S1 and S5 consume no agent tokens.

The curve relates this budget to the speedup it buys, and it exhibits the diminishing
return that a residual-driven loop predicts. The early streaming and stride engines are
localized cheaply and deliver most of the gain, whereas the sparse and irregular
residual that remains late in the schedule, together with the two new-pattern engines of
Section~\ref{sec:newpattern}, costs more per unit speedup and spends about a quarter of
the cumulative tokens for the final 0.8 points. This is the design-time analog of the
coverage result of
Section~\ref{sec:eval-cat}, since the marginal engine costs more to discover precisely
because the easy patterns are already covered. We therefore regard the \$41.66 as an
acceptable design-time budget, since it is paid once and offline before tape-out and is
amortized over the deployed lifetime of the fixed 110\,KB datapath of
Section~\ref{sec:eval-storage}.

%% file: src/conclusion.tex
\section{Conclusion}
\label{sec:concl}

We presented a performance-anomaly-driven autoresearch flow that closes specialized
agents around a deployed prefetcher. The flow diagnoses measured residual misses,
validates them with runnable minimal cases, and accumulates generated specialists
through residual gating. Rather than optimizing within a human-specified design space,
the loop turns the deployed design's remaining failures into new research questions
and uses experimental outcomes to decide which repairs survive. The resulting
17-engine \alectofinal achieves a 61.1\% held-out geomean IPC speedup over no
prefetching, outperforming Alecto, Berti, and Pythia by 14.5, 21.6, and 23.6
percentage points, while using 110\,KB and 0.0347\,mm$^2$. These results show that
agents can automate practical prefetcher improvement under simulator feedback.